# Multi-state and non-volatile control of graphene conductivity with surface electric fields

V. Iurchuk,[1] H. Majjad,[1] F. Chevrier,[1] D. Kundys,[2] B. Leconte,[1] B. Doudin,[1] and B. Kundys[1]
[1]*Institut de Physique et Chimie des Matériaux de Strasbourg (IPCMS), UMR 7504 CNRS-UdS 23 rue du Loess, 67034 Strasbourg, France*
[2]*School of Physics and Astronomy, University of Manchester, Manchester M13 9PL, Lancs, England*

Planar electrodes patterned on a ferroelectric substrate are shown to provide lateral control of the conductive state of a two-terminal graphene stripe. A multi-level and on-demand memory control of the graphene resistance state is demonstrated under low sub-coercive electric fields, with a susceptibility exceeding by more than two orders of magnitude those reported in a vertical gating geometry. Our example of reversible and low-power lateral control over 11 memory states in the graphene conductivity illustrates the possibility of multimemory and multifunctional applications, as top and bottom inputs remain accessible.
http://dx.doi.org/10.1063/1.4934738

The single-atom layer structure and the specific electronic properties of graphene provide unique opportunities for innovative devices control.[1] In particular, the high sensitivity of graphene conductivity to the presence of nearby electric charges makes the construction of ferroelectric (FE)-graphene hybrids of high interest, especially by taking advantage of the surface charges of a ferroelectric film for electric field gating purposes. Reported devices use either a top organic (p(VDF-TrFE))[2–6] or a bottom inorganic ferroelectric films[7–9] as a gating dielectric layer. The graphene conductivity can be modified by several hundred of percents[2,3] when the applied gate voltage switches the ferroelectric state of the substrate vertically. Another interesting feature in such a structure is the resistive readout of the ferroelectric-induced state, which makes the device operation comparable with the resistive random access memories (RRAMs, memristors),[10,11] where large electroresistive effects are highly desirable. Although a storage density would greatly benefit from multistate operation, the previous reports on the ferroelectric-controlled graphene structures show a bi-state performance, related to the two polarization states of the ferroelectric neighbour. This switching requires applied electric fields larger than the coercive value needed to switch the FE dipoles in the bulk, possibly prone to FE fatigue problems.

We recently demonstrated that subcoercive electric fields manipulations of bulk[12] and surface[13] configurations of ferroelectric substrates can result in stable remnant configurations, related to irreversible motions of the ferroelectric domains. This hysteretic control at subcoercive applied electric fields is very scarcely documented in the literature[14] and constitutes an alternative low power approach for both resistive and magnetoresistive based memories applications. In our previous reports, we essentially showed how a strain gage made of a metallic line can reveal the occurrence of a remnant strained state of a ferroelectric substrate, with a value tunable by the maximum stress voltage applied, as long as kept below the coercive value.[12,13] Here, we apply this emergent approach to a graphene monolayer stripe gage, to demonstrate the multistate graphene conductivity control and the improvement in the magnitude of the electric-field-induced resistance change of the device. Owing to the low graphene sensitivity to deformation,[15,16] we detail how a strain gage mechanism cannot explain the change of resistance state after the subcoercive ferroelectric voltage pulse. We rely, however, on the previous studies on the graphene–ferroelectric systems that illustrated the electrostatic nature of the origin of the sensitivity of graphene conductivity under switching of the electric polarization of the nearby ferroelectric.[2,3,5,6] The resulting modification of the charge doping of the graphene impacts significantly the film conductivity, owing to the high mobility of graphene. Unlike previous studies, here, we show a multistate operation in a lateral configuration and at sub-coercive electric fields that provide opportunities for increased device sensitivity and lower power operation. Our proposed configuration can be assimilated to a side-gating field-effect, which can provide advantages in terms of multifunctionality. Indeed, the graphene layer remains accessible from the top for exciting the samples with light or another electric field, or for the other environmental or chemical modifications of the graphene device conductivity.

The design of the fabricated sample is illustrated in Figure 1(a), with the sketch of the graphene monolayer stripe deposited onto the ferroelectric substrate with metallic side electrodes in a lateral geometry. Commercially available ceramics of lead zirconium titanate (PZT),[17] determined to be near its morphotropic phase boundary (Pb[$Zr_{0.5}Ti_{0.5}$]$O_3$), was used. The PZT ceramics of 0.5 mm thickness was received in a spontaneous ferroelectric state with a net remnant polarization perpendicular to the surface. Surface roughness was improved by polishing, down to ∼10 nm rms roughness, without post-annealing. The graphene monolayer was deposited by the CVD transfer on the top of the FE substrate. A shadow mask was used to protect the graphene monolayer stripe while the excess of the graphene was removed by the $Ar^+$ ion etching, using our established recipe known to impact the graphene transport properties only marginally.[18] The final dimensions of the graphene stripe were

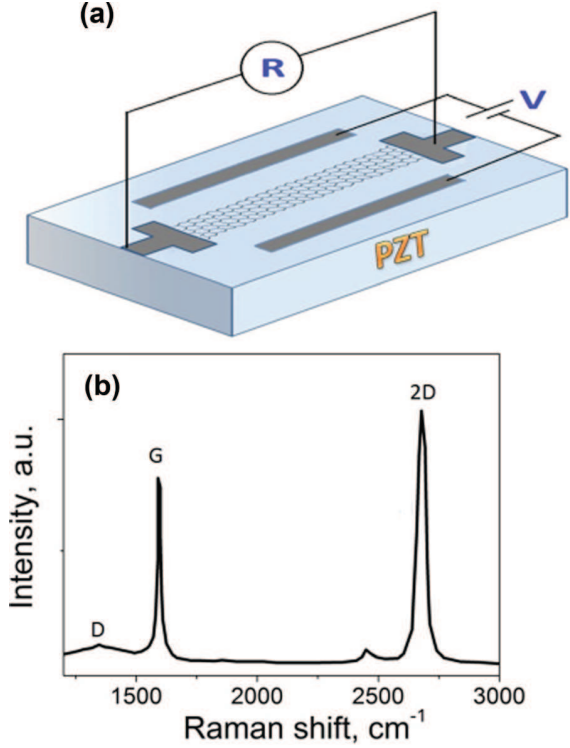

FIG. 1. The schematics of the experiment (a) and Raman spectrum of the graphene overlayer (b) measured on the same batch (on Si/SiO$_2$ (300 nm)).

240 $\mu$m in width and 1.6 mm in length. The electrodes (parallel to the stripe, Figure 1(a)) were made of Ti (10 nm) and Au (60 nm) deposited in the e-beam evaporator at a base pressure of $10^{-7}$ mbar through a stencil shadow mask. The Raman spectrum of transferred CVD graphene shown in Figure 1(b) confirms the presence of good quality single layer sample (slight D peak observed at 1350 cm$^{-1}$). More detailed studies can be found in Ref. 18. The electrodes for resistance measurements were connected using silver paste and the resistance was measured using the LCR meter at 100 kHz at ambient conditions. The initial sheet resistance of the stripe is of the order of 1 k$\Omega$/□, which is expected for our strongly p-doped layer.

In-plane electric fields applied to the surface of the ferroelectric substrate in a direction transverse to the graphene stripe (Figure 1(a)) result in a large resistive change (Figure 2) of 45%, corresponding to 420 $\Omega$/□, observed at moderate electric field of 200 kV/m. The relative change of resistance under applied electric field, or relative electroresistance susceptibility, of the order of 0.2% per kV/m, can be considered as a reasonable figure of merit of the device sensitivity to an electric field. The lateral geometry we propose has much larger amplitude in terms of susceptibility than the vertical geometry reported in the literature. For example, an electric field change of 5 MV/m is needed to obtain $\sim$64.5 $\Omega$ change only in Ref. 2, related to a susceptibility two orders of magnitude smaller than our reported values. We also recall here that we did not reach the coercive electric field region, where larger susceptibility is expected. From a technical point of view, the sub-coercive approach is expected to relate to the lower power substrate excitation and is highly beneficial for avoiding fatigue issue to the ferroelectric element.

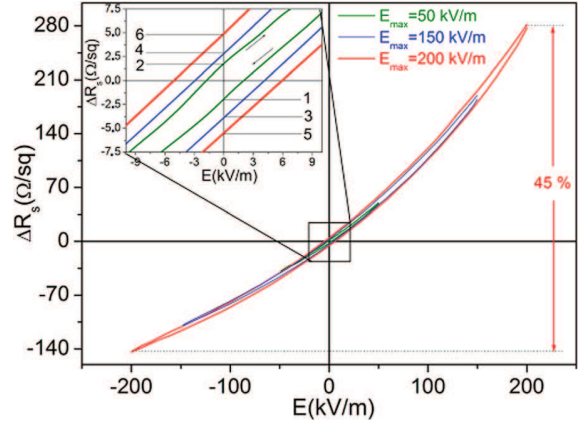

FIG. 2. Electroresistive hysteresis of the graphene stripe with the zoomed low-field behavior, showing that the zero-field remnant states, numbered 1–6, depend on the amplitude of the maximum applied field, illustrating the possibility of realizing a multi memory states device. The time constant was 1.4 s.

Alternatively, one can also compare the remnant resistance change after applying a given voltage stress value. The best result of Ref. 2 shows $\sim$3.5 k$\Omega$ variation after 120 MV/m applied field, which is $\sim$2.9 $\times$ $10^{-5}$ $\Omega$ m/V resistance susceptibility. This is more than one order lower than our values ($\sim$71.5 $\times$ $10^{-5}$ $\Omega$ m/V) observed on the samples for non-optimized structure. The resistance vs electric field loops are asymmetrical and show a hysteresis with remnant resistive states (Figure 2, inset). For a sample in its initial state with resistance R0, other remnant or memory resistance states are observed when the electric field is returned back to zero. By increasing the maximum applied electric field, one can form multiple memory states (numbered 1–6 in Figure 2, inset). To understand the possible origin of the observed electroresistive hysteresis, two mechanisms should be considered: (i) the electric field induced deformation of the surface that is transferred to graphene and (ii) a surface charge change resulting in a modified graphene doping.

Taking into account the applied electric field values, the expected deformation is unlikely to exceed 0.03% (Ref. 12) at our maximum applied electric field range of 200 kV/m. The reported low susceptibility of the graphene conductivity to deformations[14,15,19,20] should result in a negligible resistance change, of the order of our measurement noise, and therefore excludes a strain-dominated mechanism. The substrate doping mechanism is also debatable because there should be no free charges in an ideal ferroelectric under the subcoercive electric field stress. The possible types of charge movements contributing to electric currents in ferroelectrics are dipoles orientation and displacement currents. The first one is larger, as it originates from dipole movement under large electric fields of the coercive value. The second one is smaller, as it results from bound charge displacement from their average positions: domains at the sub-coercive fields and atoms at above the coercive field region. Due to its secondary role and much smaller contribution, the displacement current is usually neglected in ferroelectric systems. It is nevertheless of great importance to study the displacement current dynamics as it can be used for multilevel data storage operating at low electric fields.[21,22] We therefore measured the surface electric current between the two side electrodes (Figure 1) *in-situ* when applying the voltage resulting in the data of Figure 2. Indeed, the displacement current dynamics shows a well-defined hysteresis behavior (Figure 3(a)). The initial zero electric current indicates the measurements free from parasitic capacitance in the circuit. The time integrated current provides the charge hysteresis cycle shown in Figure 3(b), which exhibits a slightly larger charge absolute values for positive electric field, thus reproducing the asymmetry of the electroresistive behavior shown in Figure 2. It therefore confirms that the graphene resistance change relates to the modification of doping, caused by the surface charge movements of the ferroelectric substrate induced by the lateral electric field. As the electric field is applied (in our conventional positive direction in Figures 2 and 3), the charges concentrated at the domain walls start to separate and move towards negative and positive electrodes, respectively. Since our graphene is in the form of a stripe and situated in the middle of the structure, the charges leave the stripe area, decrease the p-doping of the graphene, leading to a resistance increase. When the electric field starts to decrease, the charge re-distribution leads to the resistance decrease with a remnant state "1" smaller than "0," in agreement with the presence of the larger change amount (Figure 3(b)). The remanence effects therefore relate to a modification of the ferroelectric domains configurations, perturbed by the voltage stress history experienced by the sample. This is illustrated in Figure 4(a) (insets), showing how the configuration of the ferroelectric domains at zero applied field depends on the electric history of the sample. Figure 4(a) confirms the stability and diversity of the multiple resistance states created by this surface geometry approach.

The number of presented resistance state values is only limited here by the resolution in the resistance measurements and sub-coercive electric field region. Although we cannot guarantee experimentally (as sub-100 nm ferroelectric domains imaging are necessary) that the exact domain configuration of the virgin state can be reproduced, the corresponding surface charge equilibrium and therefore the corresponding initial resistance of the graphene can be recovered on-demand by applying a damping voltage procedure, with a decreasing and oscillating time profile shown in Figure 4(b). Therefore, an additional "erased" memory state R0 of the graphene stripe can be formed and recovered on-demand. Note that this "erasing" procedure was applied before each loop, shown in Fig. 2, to erase the poling history. The multiplicity of the possible remnant states most likely relates to the multiplicity of the domain wall pinnings in the FE domains structure. For example, carrier density modulation in graphene was recently achieved by spatially defined potential steps at 180° domain walls in Ref. 23. Scaling down the size of the device allows diminishing the applied voltage range below 1 V, as expected from the low electric field values we use. One can also expect that sub-100 nm domains structure can be stabilized in thin films and controlled in micron-sized cells,[24,25] making a multi-state approach a way forward to circumvent the difficulties in stabilizing the polarization states of the FE cells when miniaturizing the device.[26] Reducing the size of the sample diminishes accordingly the applied side voltage, easily reaching the sub-1V range for devices in the $\mu$m size range. It is also expected to increase the difference in the remnant resistance state values, as our presented large-scale resistance memory states involve the averaging of many domains configurations. Our initial tests on the switching and lifetime durability of the graphene conductivity states revealed up to $10^6$ cycles, where the difference R0-RN remains reproducible without any sign of fatigue for 7 days. Miniaturization may provide easier implementation of faster pulses excitations, with expected high sample resistance to fatigue, as we avoid the coercive operation of the FE substrate, known to be the main durability obstacle in these materials.

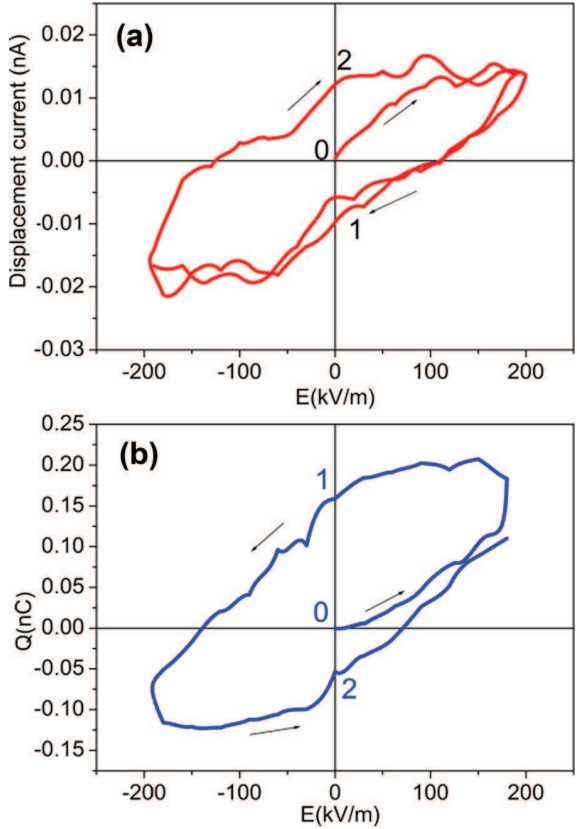

FIG. 3. Electric current between the two electrodes creating an electric field perpendicular to the graphene stripe (Figure 1) (a) and corresponding surface charge hysteresis obtained by the current integration with respect to time method (b). The time constant was 1.4 s.

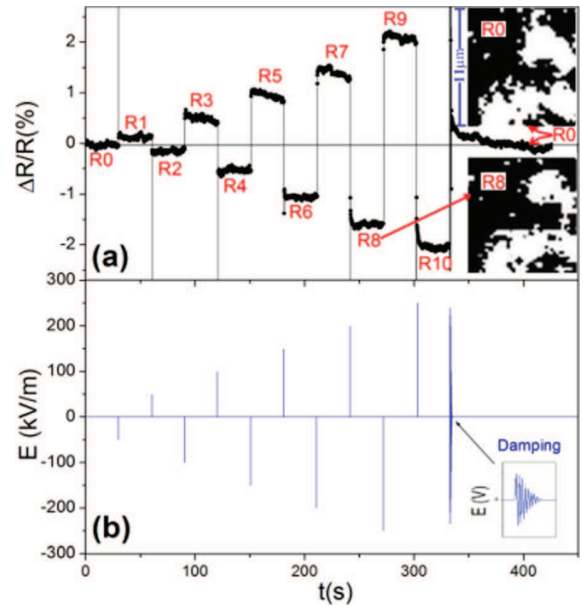

FIG. 4. Electric field induced switching between multiple memory states (a) with a corresponding electric field pulses (of 100 ms duration) time profile (b). The insets (a) show an optical micrograph under polarized light, revealing the differences in ferroelectric domains configurations.

Our proof-of-principle device design and operation provide, therefore, compelling evidence that surface electric fields' control of ferroelectric substrate paves an opportunity to manipulate the graphene resistance at lower electric fields and with high efficiency. The governing mechanism is the surface charge dynamics of the charged domains induced by the lateral electric field, which fascinatingly results in multistate operation possibility. The magnitude of the electroresistance effect itself can be greatly improved by gating the reported structure with a proper charge mobile environment. Importantly for application, we have also shown that any of the stored memory sates can be easily erased on demand by applying an oscillating damped voltage pulse. More generally, the lateral geometry in ferroelectrics is highly desirable for planar 2D cell design and can offer a unique opportunity to tune the properties of a surface via an electric field, potentially providing an avenue for some additional advanced surface functionalities.[27] Considering the importance of graphene for future types of electronic and sensor devices, our results may provide an opportunity for additional functionalities of graphene-based devices operating at nonvolatile and multistate principle with lower and surface electric fields. The different graphene charge doped states can in principle be used to perform electric field manipulations of band gap opening in the graphene bi-layer structures[28] or efficiency improvements in the graphene-based solar cells.[29] Our results may also contribute to the possibility of creating a tunable graphene layers for spin injection in a spintronic device[30–32] or even for ferroelectric tunnel junctions with the graphene electrodes[33] with enhanced low power and multistate storage advantages.

Corresponding author email: kundys(A)ipcms.unistra.fr